\documentclass[sigplan,screen,nonacm]{acmart}
\renewcommand\footnotetextcopyrightpermission[1]{}
\microtypesetup{expansion=false}

\AtBeginDocument{%
  }

\usepackage{amsmath}
\usepackage{filecontents}
\usepackage[normalem]{ulem}
\usepackage{subcaption}
\usepackage{graphicx}
\setkeys{Gin}{draft=false}
\graphicspath{{pdfs/}}
\usepackage{filecontents}
\usepackage{enumitem}
\usepackage{pifont}
\usepackage{booktabs}

\usepackage{cleveref}
\crefname{section}{§}{§§}
\Crefname{section}{§}{§§}
\usepackage{bm}
\usepackage{array}
\usepackage{wrapfig}
\usepackage{balance}
\usepackage{multirow}
\usepackage{caption}
\usepackage{threeparttable}
\usepackage{todonotes}
\usepackage{diagbox}
\usepackage{wrapfig}
\usepackage[most]{tcolorbox}    
\usepackage{enumitem}
\usepackage{xspace}

\renewcommand{\emph}[1]{\textit{#1}}
\newcommand{\sys}{CodecSight\xspace} %
\ifdefined\RELEASE
  \newcommand{\ignore}[1]{}
  \newcommand{\fixme}[1]{}
  \newcommand{\dmi}[1]{}
  \newcommand{\yu}[1]{}
  \newcommand{\wy}[1]{}
  \newcommand{\yan}[1]{}
  \newcommand{\franky}[1]{}
  \newcommand{\TODO}[1]{}
  \newcommand{\col}[1]{}

\else
  \newcommand{\ignore}[1]{}
  \newcommand{\fixme}[1]{{\textcolor{red}{[~FIXME:~#1~]}}}
  \newcommand{\dmi}[1]{{\textcolor{blue}{[~D:~#1~]}}}

  \newcommand{\yu}[1]{{\textcolor{teal}{[~Yu:~#1~]}}}
  \newcommand{\wy}[1]{{\textcolor{olive}{[~W:~#1~]}}}
  \newcommand{\yan}[1]{{\textcolor{gray}{[~Ya:~#1~]}}}
  \newcommand{\franky}[1]{{\textcolor{brown}{[~F:~#1~]}}}
  \newcommand{\TODO}[1]{{\textcolor{red}{TODO:~#1}}}
  \newcommand{\col}[1]{{\textcolor{pink}{[~Co：~#1~]}}}
  
\fi

\begin{document}
\sloppypar
\widowpenalty=1000
\clubpenalty=1000
\brokenpenalty=1000

\title{\sys: Leveraging Video Codec Signals for Efficient Streaming VLM Inference}

\author{Yulin Zou}
\affiliation{\institution{NTU Singapore}\country{Singapore}}

\author{Wenyan Chen}
\affiliation{\institution{NTU Singapore}\country{Singapore}}

\author{Yan Chen}
\authornote{Work done while at NTU Singapore.}
\affiliation{\institution{Beihang University}\country{China}}

\author{Anya Rajan}
\affiliation{\institution{NTU Singapore}\country{Singapore}}

\author{JooYoung Park}
\affiliation{\institution{NTU Singapore}\country{Singapore}}

\author{Shivaraman Nitin}
\affiliation{\institution{A*STAR IAIC}\country{Singapore}}

\author{Luo Tao}
\affiliation{\institution{A*STAR IAIC}\country{Singapore}}

\author{Francisco Romero}
\affiliation{\institution{Georgia Institute of Technology}\country{USA}}

\author{Dmitrii Ustiugov}
\affiliation{\institution{NTU Singapore}\country{Singapore}}

\renewcommand{\shortauthors}{Zou et al.}

\begin{abstract}
Continuous inference over concurrent video streams imposes substantial compute and memory demands on vision-language model (VLM) serving. Streaming inference uses sliding windows to maintain a bounded context of recent video, but processing each window independently repeats visual encoding and large language model (LLM) prefilling for similar and overlapping content. Existing optimizations provide limited coordination across these stages and often rely on model-specific training, profiling, or model-generated signals.

We present \textit{\textbf{\sys}}, a streaming VLM serving system that uses codec metadata as shared runtime guidance across visual encoding and LLM prefilling, without model-specific training or offline profiling. Codec-derived change signals guide patch pruning before visual encoding, reducing both visual computation and the number of downstream visual tokens. Codec-defined frame types guide selective key--value (KV) refresh across windows, while positional correction enables reuse of the remaining cached keys. Across three VLMs and four video workloads, our vLLM-based implementation supports up to $3.3\times$ as many concurrent streams and achieves up to a $5.3\times$ speedup in average time-to-first-token relative to the state-of-the-art baselines. It also reduces executed FLOPs by up to 93\%, with a maximum task-quality decrease of 4.64 percentage points.
\end{abstract}

\maketitle

\section{Introduction}
\label{sec:intro}

Streaming video understanding is emerging as an important serving workload, requiring systems to interpret incoming video and answer queries as a stream evolves~\cite{liu2024streamchat,chen2025livecc,di2025streaming,niu2025ovo}. Applications such as traffic monitoring~\cite{zhou2025vision,sharma2026scaling}, retail analytics~\cite{ou2025real}, and industrial monitoring~\cite{kurrey2025process} require this ongoing analysis to track events and changes over time. The potential deployment scale is substantial: surveillance cameras alone are estimated to exceed 1.1 billion worldwide~\cite{persistence2024cctv,marketreportsworld2026cctv}. Realizing this potential requires video understanding systems that can serve diverse analysis tasks across many concurrent streams. Vision-language models (VLMs)~\cite{li2024llavaov,bai2025qwen3,chen2024internvl} provide a general foundation for these tasks by jointly reasoning over visual content and language queries~\cite{wu2025monitorvlm}.

\begin{figure}[t]
    \centerline{\includegraphics[width=1\linewidth]{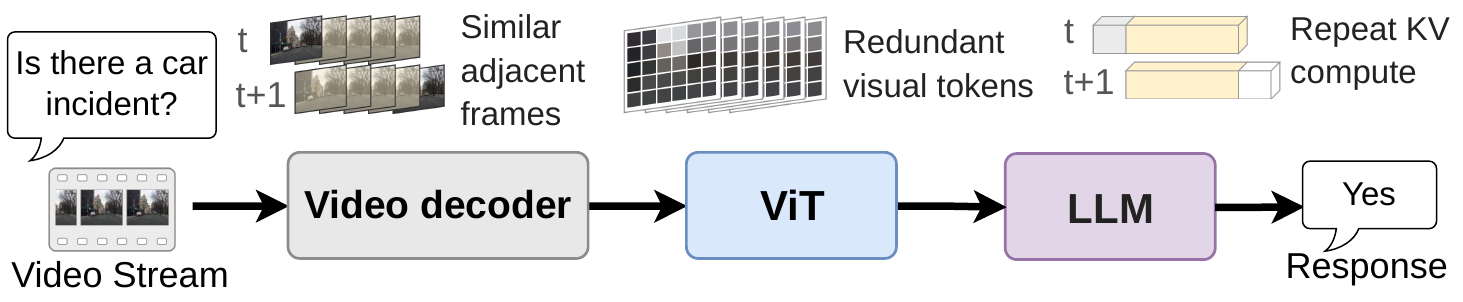}}
    \vspace{-1em}
    \caption{VLM serving pipeline over sliding windows.
    }
    \vspace{-1em}
    \label{fig:pipeline}
\end{figure}

\begin{figure}[t]
    \centering
        \begin{subfigure}[t]{0.48\linewidth}
        \centering
        \includegraphics[width=\linewidth]{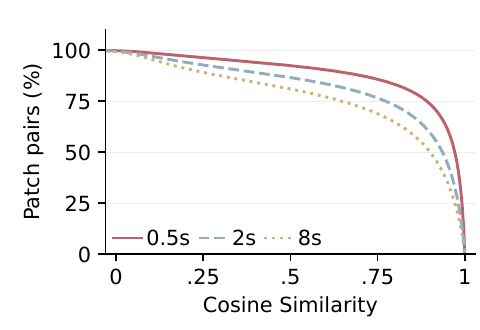}
        \vspace{-2em}
        \caption{Patch similarity CCDF}
        \label{fig:cosine_similarity}
    \end{subfigure}
    \hfill
    \begin{subfigure}[t]{0.48\linewidth}
        \centering
        \includegraphics[width=\linewidth]{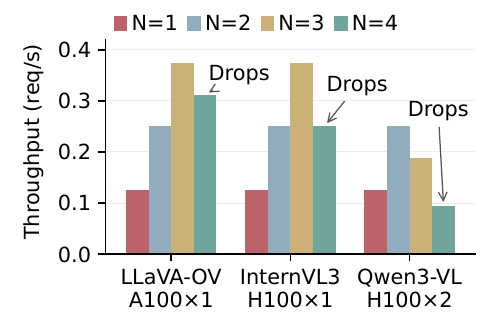}
        \vspace{-2em}
        \caption{Throughput}
        \label{fig:throughput_full_compute}
    \end{subfigure}
    \vspace{-1em}
    \caption{Temporal redundancy and concurrency scaling in streaming VLM serving. (a) Spatially corresponding patches remain highly similar across multi-second intervals. (b) Under default serving pipeline, aggregate throughput saturates or declines as concurrent streams $N$ increase.}
    \label{fig:serving_bottlenecks}
    \vspace{-1.2em}
\end{figure}

To answer queries over continuously arriving video, streaming VLM inference commonly processes bounded sliding windows that advance over time~\cite{shen2026simple}. Fig.~\ref{fig:pipeline} illustrates the conventional VLM serving pipeline: a video decoder reconstructs frames, a visual encoder (ViT) converts sampled frames into visual tokens, and an LLM prefills these tokens with a text query to construct key--value (KV) states before generating a response. This execution pattern exposes \textbf{substantial temporal redundancy} at two levels. \emph{Within each window}, frames can remain highly similar even across multi-second intervals. In our measurements at 2\,FPS on UCF-Crime~\cite{sultani2018real}, we compare $14$$\times$$14$-pixel patches at matching spatial locations between sampled frames separated by different temporal intervals. Fig.~\ref{fig:cosine_similarity} shows that even at an 8\,s interval, approximately 70\% of corresponding patch pairs have cosine similarity above 0.75. \emph{Across consecutive windows}, overlapping frames undergo repeated visual encoding and LLM prefilling. These per-stream costs compound under concurrency, limiting the scalability of the default serving pipeline: Fig.~\ref{fig:throughput_full_compute} shows that aggregate throughput saturates or declines at only 3--4 concurrent streams for the three VLMs under their respective GPU configurations.

This limited throughput stems from repeatedly processing overlapping and temporally redundant visual content through both the ViT and LLM. Yet, aggressive skipping risks discarding important visual evidence, as redundant-looking frames may contain subtle query-relevant changes. The system must therefore reduce repeated computation across stages while preserving such evidence. This motivates a critical question: \emph{How can streaming VLM serving coordinate redundancy reduction across visual encoding and LLM prefilling while preserving inference quality?}

\begin{figure}[t]
    \centerline{\includegraphics[width=1\linewidth]{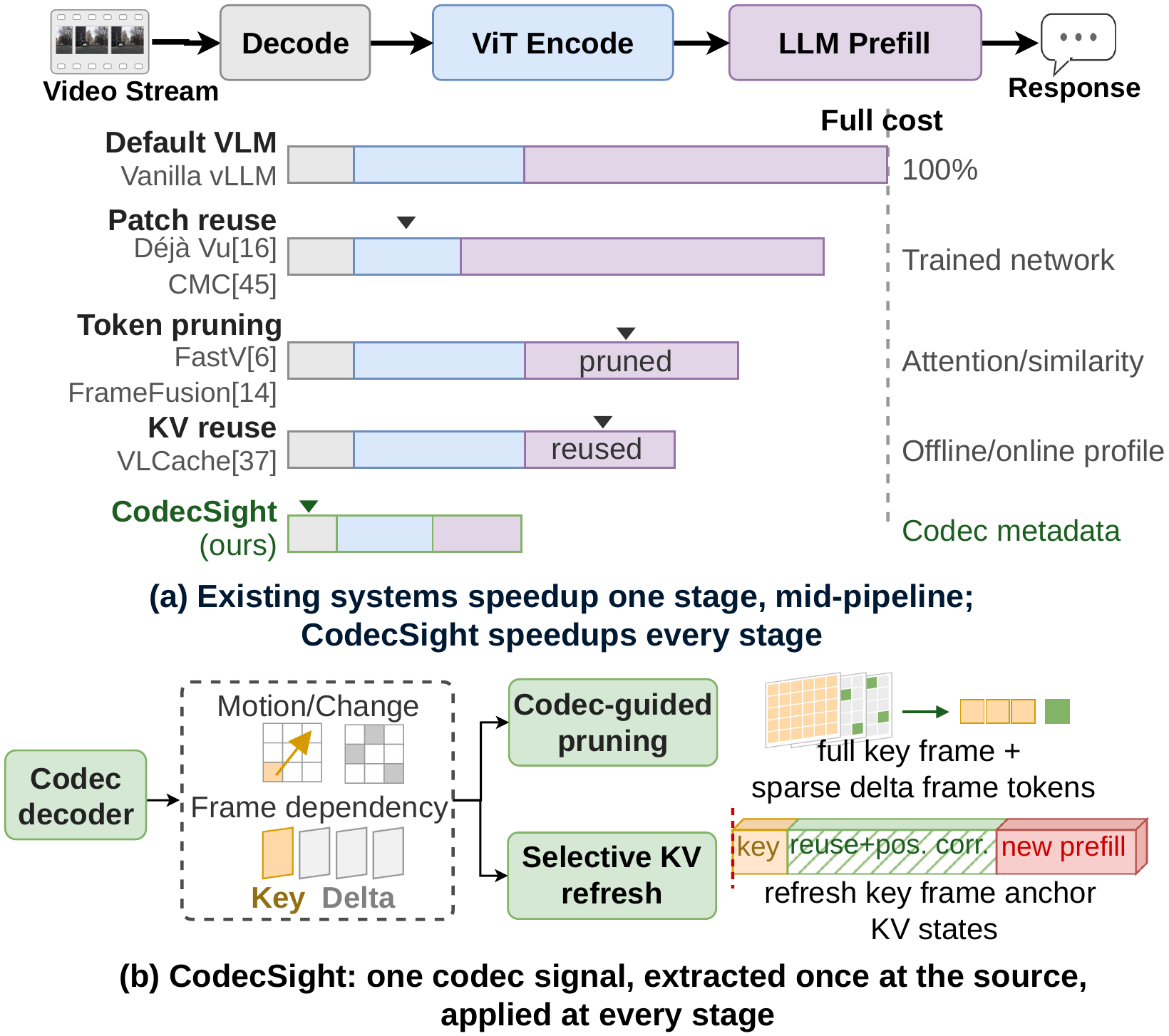}}
    \vspace{-1em}
    \caption{The novel mechanisms in \sys that differ from existing systems.
    }
    \label{fig:costream}
    \vspace{-1em}
\end{figure}

\textbf{Existing Systems.}
Recent systems reduce redundancy at different stages of streaming VLM serving through patch reuse, token pruning, and KV/state reuse~\cite{hwang2025dejavu,song2024cmc,chen2024image,fu2025framefusion,qin2025vlcache,wu2026metom}. However, two limitations restrict their end-to-end benefits in streaming serving (Fig.~\ref{fig:costream}(a)). \textbf{\textit{L1: Costly redundancy decisions.}} Many pruning and reuse policies rely on model-specific training or profiling~\cite{hwang2025dejavu,qin2025vlcache}, or on features, similarity, or attention signals produced during model execution~\cite{chen2024image,fu2025framefusion}, incurring preparation or runtime overhead. \textbf{\textit{L2: Limited cross-stage coordination.}} Existing approaches decide on stage-specific representations, such as visual patches~\cite{hwang2025dejavu,song2024cmc} or LLM KV states~\cite{qin2025vlcache}. Simply combining these optimizations does not ensure that their decisions remain compatible across windows. For example, visual selection based on the current window can retain different tokens for the same overlapping frame as its context or reference frames change, complicating direct reuse of its cached KV states. Together, these limitations motivate lightweight source-side signals that coordinate visual reduction and cross-window state reuse without model-specific training or profiling.

\textbf{Insights and Challenges.}
Our key insight is to use source-side signals available before model execution as shared runtime guidance for redundancy reduction across the decode--ViT--LLM pipeline. Codec metadata~\cite{wiegand2003overview,richardson2024coding} provides such signals: motion vectors and residuals capture block-level displacement and prediction error that can guide early patch selection; frame types and GOP (group of pictures) boundaries provide a basis for organizing frame-aligned token spans as candidates for KV refresh and reuse across windows. Realizing this opportunity, however, introduces two key challenges. \textbf{\textit{C1: Codec-to-model representation gap.}} Codec metadata is defined over coding blocks and prediction dependencies, whereas VLMs operate on resized patch grids and visual tokens; these signals must be mapped to the model's patch layout and translated into sparse execution without discarding important visual evidence. \textbf{\textit{C2: Context-sensitive state reuse.}} Visual pruning changes the token sequence on which KV reuse operates, potentially altering both the correspondence with cached states and the preceding context of retained tokens. The challenge is to use codec cues to coordinate pruning with selective refresh across windows, retaining as much cached computation as possible while limiting reuse error.

\textbf{\sys.}
To address these challenges, we present \textit{\sys}, a codec-guided streaming VLM serving system that turns codec metadata into cross-stage runtime control from video decoding to VLM execution (Fig.~\ref{fig:costream}(b)). \sys extracts motion and frame/GOP information once along the hardware-accelerated decode path and tracks the mapping from source frames to visual tokens and cached KV spans across overlapping windows. To address \textbf{C1}, \sys introduces codec-guided \textbf{selective visual encoding}: it maps codec-derived visual-change signals onto each VLM's patch layout and selects patches before ViT encoding, reducing both visual computation and visual tokens entering LLM prefilling. To address \textbf{C2}, \sys introduces codec-guided \textbf{selective KV refresh}: it uses frame types and GOP boundaries to define token-level refresh and reuse spans, recomputes selected anchor KV states under the updated context, and applies positional correction to reused keys. \sys further integrates state-selective refresh into the serving execution path, coupling refresh computation and new-token prefill so that they can be jointly scheduled across streams. Together, these mechanisms use shared source-side metadata to coordinate visual encoding and cross-window KV reuse without model-specific training or offline profiling.

We implement \sys\footnote{We will release the code upon publication.} on top of vLLM~\cite{kwon2023efficient} and evaluate it across three representative VLMs and four video understanding workloads, against Full Compute and prior systems including D\'ej\`a Vu~\cite{hwang2025dejavu}, CacheBlend~\cite{yao2025cacheblend}, and VLCache~\cite{qin2025vlcache}. \sys improves mean time-to-first-token by up to 5.3$\times$ and supports up to 3.3$\times$ as many concurrent streams with a maximum task-quality drop of 4.64 percentage points across all tasks. It reduces executed FLOPs and new KV volume by up to 93\% and 90\%, respectively.

\noindent\textbf{Contributions.} In summary, our main contributions are:
\begin{itemize}[leftmargin=0cm,itemindent=.3cm,labelwidth=\itemindent,labelsep=0cm,align=left]

    \item \textit{Cross-stage codec-to-model runtime control.}
    We introduce a runtime abstraction that uses codec-derived signals to coordinate sparse visual encoding and cross-window KV refresh and reuse across the decode--ViT--LLM boundary.

    \item \textit{Codec-guided visual encoding and KV refresh.}
    We translate codec motion into pre-ViT patch selection and frame/GOP structure into token-level refresh and reuse spans across overlapping windows.

    \item \textit{End-to-end evaluation.}
    We implement \sys on vLLM and evaluate it across various workloads under concurrent multi-stream serving, demonstrating substantial gains in throughput, latency, computation, and new KV volume alongside task-quality changes.
\end{itemize}

\vspace{-.5em}
\section{Background and Motivation}
\label{sec:background}

\subsection{Streaming VLM Serving Pipeline}
\label{sec:vlm_pipeline}

As shown in Fig.~\ref{fig:pipeline}, a representative streaming VLM serving pipeline processes each video request through decoding, visual encoding, and LLM inference. Given a compressed video stream, the system first decodes the bitstream into frames and applies model-specific preprocessing. A visual encoder (ViT)~\cite{dosovitskiy2020image} then processes sampled frames into visual embeddings, which are projected into visual tokens. The LLM subsequently processes the visual and text tokens during prefill, caching their key--value states for autoregressive response generation.

Streaming inference commonly operates on fixed-duration \textit{sliding windows} that advance by a shorter stride as new video arrives. This execution pattern introduces temporal \textit{redundancy} at two levels: Within window, nearby frames often contain highly similar visual content; across consecutive windows, overlapping frames that have already been processed reappear in later requests. Both forms of redundancy cause successive requests to repeatedly process similar or previously observed content, increasing the computation required to keep up with each stream.

\vspace{-.5em}
\subsection{Scaling Challenges in Streaming VLM Serving}
\label{sec:system_imbalance}
Repeated computation within each stream limits the scalability of concurrent VLM serving. With a representative 40\,s window and an 8\,s stride, consecutive windows overlap by 80\%. Although the overlapping frames are unchanged, their token positions and preceding context change as the window advances, yielding different KV states for the current window. Full Compute consequently processes the full 40\,s of video for every 8\,s that arrives. Moreover, frames within a window can contain highly similar patches across multi-second intervals (Fig.~\ref{fig:cosine_similarity}), causing the model to repeatedly process similar visual content. 

\textbf{Concurrency amplifies redundancy cost.} 
To quantify the serving impact, we profile the \textit{Full Compute} pipeline under increasing stream concurrency on LLaVA-OneVision (1$\times$A100), InternVL3-14B (1$\times$H100), and Qwen3-VL-32B (2$\times$H100) with dataset UCF-Crime~\cite{sultani2018real}. Built on vLLM, \textit{Full Compute} encodes all sampled frames and computes the corresponding visual-token KV states in every window independently. With an 8\,s stride, each stream generates \(0.125\) requests per second, yielding an aggregate offered load of \(0.125N\) req/s for \(N\) streams. Sustainable serving requires throughput to keep pace with this load; otherwise, pending windows accumulate and results become increasingly stale. As Fig.~\ref{fig:throughput_full_compute} shows, throughput quickly saturates or declines as concurrency increases. At four streams, all three model-device configurations deliver less than \(0.35\) req/s, below the \(0.5\) req/s needed to keep pace with incoming video.

\begin{figure}[t]
    \centerline{\includegraphics[width=.9\linewidth]{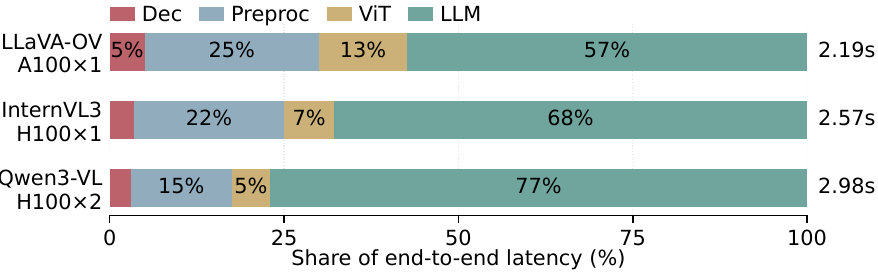}}
    \vspace{-1.1em}
    \caption{Latency breakdown under different models and configurations.}
    \label{fig:latency_breakdown}
    \vspace{-1em}
\end{figure}

\textbf{Redundancy propagates across stages.}
To understand the processing costs underlying this capacity limit, we examine the per-window latency breakdown. As shown in Fig.~\ref{fig:latency_breakdown}, total latency ranges from 2.19 to 2.98\,s, with LLM prefilling accounting for 57$\sim$77\% of the total. Although ViT encoding contributes a smaller fraction, its outputs determine the visual-token sequence subsequently processed by the LLM. Even with InternVL3's native \(4\times\) visual-token reduction, each \(448\times448\) frame produces 256 visual tokens. At 2\,FPS, a 40\,s window therefore contains 20,480 visual tokens, 80\% of which originate from frames already present in the preceding window. Together with the temporal similarity observed in Fig.~\ref{fig:cosine_similarity}, these measurements highlight the cost of processing redundant visual content in both ViT encoding and the dominant LLM-prefill stage, motivating coordinated redundancy reduction across both stages.

\vspace{-.5em}
\subsection{Limitations of Existing VLM Optimizations}
\label{sec:limitations}

\begin{table}[t]
\centering
\caption{Optimization scope and decision overhead.
Online scoring uses model representations.}
\vspace{-1em}
\renewcommand{\arraystretch}{0.8}
\resizebox{\linewidth}{!}{
\begin{tabular}{c|cc|cc}
\toprule
\multirow{2}{*}{\textbf{Method}} &
\multicolumn{2}{c|}{\textbf{Optimization Scope}} &
\multicolumn{2}{c}{\textbf{Decision Overhead}} \\
& \textbf{ViT} & \textbf{LLM}
& \textbf{Offline Prep.} & \textbf{Online Scoring} \\
\midrule
Default VLM
    & $\times$ & $\times$ & -- & -- \\
Déjà Vu~\cite{hwang2025dejavu}
    & $\checkmark$ & $\times$ & Training & Trained Predictor \\
FastV~\cite{chen2024image}
    & $\times$ & $\checkmark$ & -- & Attention \\
FrameFusion~\cite{fu2025framefusion}
    & $\times$ & $\checkmark$ & Calibration & Similarity + Attention \\
VLCache~\cite{qin2025vlcache}
    & $\checkmark$ & $\checkmark$ & Profiling & -- \\
\midrule
\textbf{\sys (Ours)}
    & $\checkmark$ & $\checkmark$ & -- & -- \\
\bottomrule
\end{tabular}
}
\vspace{-1.3em}
\label{tab:comparison}
\end{table}

Recent systems reduce redundant computation through visual/token reduction~\cite{hwang2025dejavu,song2024cmc,chen2024image,fu2025framefusion} and feature or KV-state reuse~\cite{qin2025vlcache}. Table~\ref{tab:comparison} summarizes representative approaches by optimization scope and deployment requirements. These approaches effectively reduce computation in their target stages, yet coordination between stages remains limited.

The underlying issue is that \textit{redundancy decisions are derived independently from stage-specific representations}: visual reduction uses visual patch or feature signals, while KV reuse relies on the later LLM states. The same source of redundancy is analyzed separately for visual reduction and KV reuse. Moreover, visual reduction can save computation while changing the token sequence on which downstream KV reuse depends, potentially reducing reuse opportunities. Savings from visual reduction therefore do not automatically translate into corresponding end-to-end gains.

\textit{Deriving these decisions also introduces preparation or runtime overhead.} D\'ej\`a Vu~\cite{hwang2025dejavu} learns inter-frame reuse decisions, while VLCache's layer-adaptive policy~\cite{qin2025vlcache} profiles layer sensitivity on auxiliary data. Training-free methods such as FastV and FrameFusion derive pruning decisions from attention or feature similarity during inference~\cite{chen2024image,fu2025framefusion}, after part of the model computation has already been incurred. Together, these limitations motivate lightweight signals available before model execution to coordinate visual reduction and cross-window state reuse without model-specific training or profiling.

\vspace{-.3em}
\subsection{Cross-Stage Opportunities from Codec Metadata}
\label{sec:opportunities_and_challenges}
\begin{figure}[t]
    \vspace{-0.5em}
    \centerline{\includegraphics[width=\linewidth]{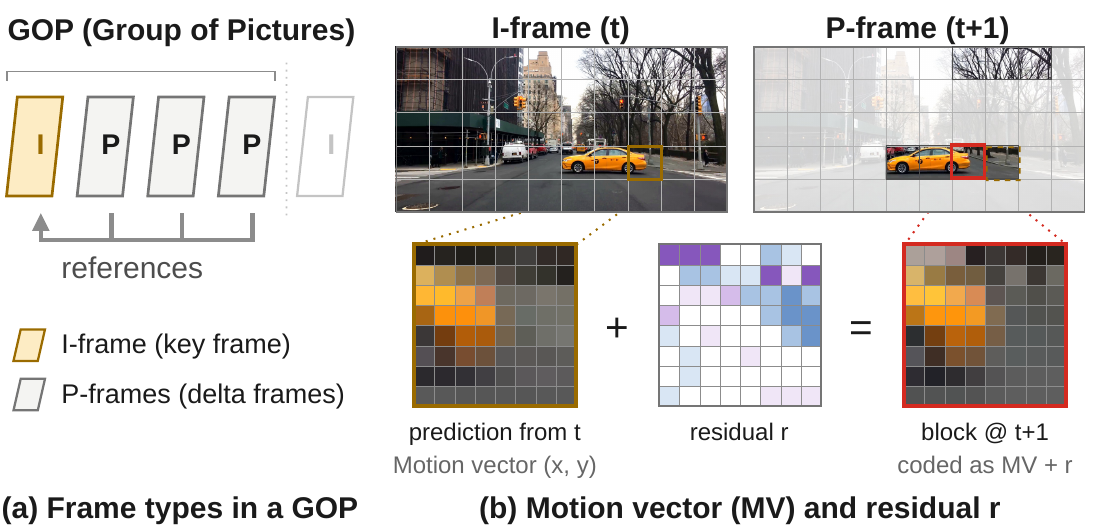}}
    \vspace{-1em}
    \caption{Codec metadata exposes block-level changes and temporal frame dependencies.}
    \vspace{-0.5em}
    \label{fig:codec_guide_example}
\end{figure}

The preceding analysis motivates a source-side signal available before VLM execution and usable across stages and overlapping windows. Video codecs~\cite{wiegand2003overview,richardson2024coding} expose such metadata along the decode path at two granularities (Fig.~\ref{fig:codec_guide_example}). At the frame level, a common GOP configuration for low-latency streaming consists of an intra-coded I-frame followed by P-frames. The I-frame is coded without reference to other frames, whereas P-frames predict image regions, called coding blocks, from previously coded reference frames. At the block level, motion vectors specify displacement to reference blocks, while residuals represent differences between current blocks and their motion-compensated predictions. %

These two granularities support two optimization opportunities. Block-level change signals can guide selective visual processing before ViT execution; the resulting visual sparsity directly shortens the token sequence entering LLM prefilling, reducing computation across both stages. Meanwhile, frame types and GOP boundaries provide a basis for organizing frame-aligned token spans as candidates for KV refresh and reuse. Importantly, codec metadata is tied directly to frames rather than logical windows: when the same frame reappears in overlapping windows, its codec signals can be reused without repeated model-side analysis. 

\begin{figure}[t]
    \vspace{-0.5em}
    \centerline{\includegraphics[width=.8\linewidth]{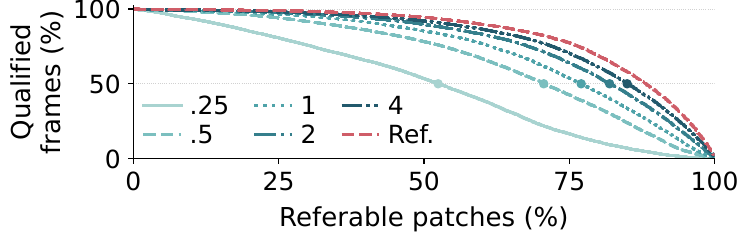}}
    \vspace{-1em}
    \caption{CCDF of per-frame referable-patch ratios under different MV-magnitude thresholds. Referable patches have a valid reference and motion below $\tau$; \textit{Ref.} shows the reference coverage without MV filtering.}
    \vspace{-1em}
    \label{fig:mv_residual_analysis_cdf}
\end{figure}

\textbf{Opportunity\#1: Propagate early visual sparsity downstream.}
Inter-frame prediction exposes spatially localized changes before ViT execution. By mapping motion vectors onto the ViT patch grid, the system can identify low-motion patches as candidates for selective visual processing. To quantify this opportunity, we analyze MVs on UCF-Crime~\cite{sultani2018real} and define a patch as \emph{referable} when its mapped motion magnitude falls below an MV threshold $\tau$. Fig.~\ref{fig:mv_residual_analysis_cdf} plots the CCDF of the per-frame referable-patch ratio for $\tau\in\{0.25,0.5,1,2,4\}$ pixels; larger thresholds classify more patches as referable. At the median, P-frames contain 52\%--85\% MV-referable patches across these thresholds, revealing substantial potential for patch pruning before ViT execution. Exploiting this opportunity can reduce ViT computation and shorten the visual-token sequence entering the LLM, reducing prefill computation and KV-cache footprint.

\textit{\textbf{Design Requirement $R_1$: Model-compatible and accuracy-preserving visual sparsity.}}
Low codec-level motion alone does not imply low importance. The system must translate block-level change signals into actual sparse execution across model-specific resizing, ViT patch layouts, and visual-token projections while preserving query-relevant evidence. This translation must also remain lightweight enough that its decision cost does not offset the computation eliminated by pruning.

\textbf{Opportunity\#2: Use codec frame-level metadata to guide selective KV refresh.}
Overlapping windows repeatedly prefill the LLM for previously observed visual content. Codec-reported I-frames and GOP boundaries provide reference points for organizing this content into frame-aligned visual-token spans. In particular, I-frame token spans can serve as refresh anchors, allowing selected states to be recomputed under the current context while retaining the remaining overlapping states for reuse. This offers a source-guided alternative to selecting refresh candidates through model-generated importance scores.

\textit{\textbf{Design Requirement $R_2$: Coordinate visual pruning with context-sensitive KV reuse.}}
Visual pruning determines which tokens enter the LLM, while window advancement changes their positions and preceding context. A refresh plan must therefore account for the retained token sequence and its correspondence with cached states. The runtime must coordinate token selection with selective recomputation and position correction, retaining as much cached computation as possible while limiting reuse error and state-management overhead.

\vspace{-.5em}
\section{System Design}
\label{sec:system_design}
We present \sys, a codec-guided serving system that uses codec metadata as shared runtime signals to coordinate visual encoding and LLM prefilling. As motivated in \cref{sec:opportunities_and_challenges}, \sys maps block-level change signals to ViT patches for model-compatible selective visual encoding, and uses frame types and GOP boundaries to guide selective KV refresh and reuse across overlapping windows.

\vspace{-.3em}
\subsection{System Overview}
\label{sec:system_overview}
\begin{figure}[t]
    \centerline{\includegraphics[width=1\linewidth]{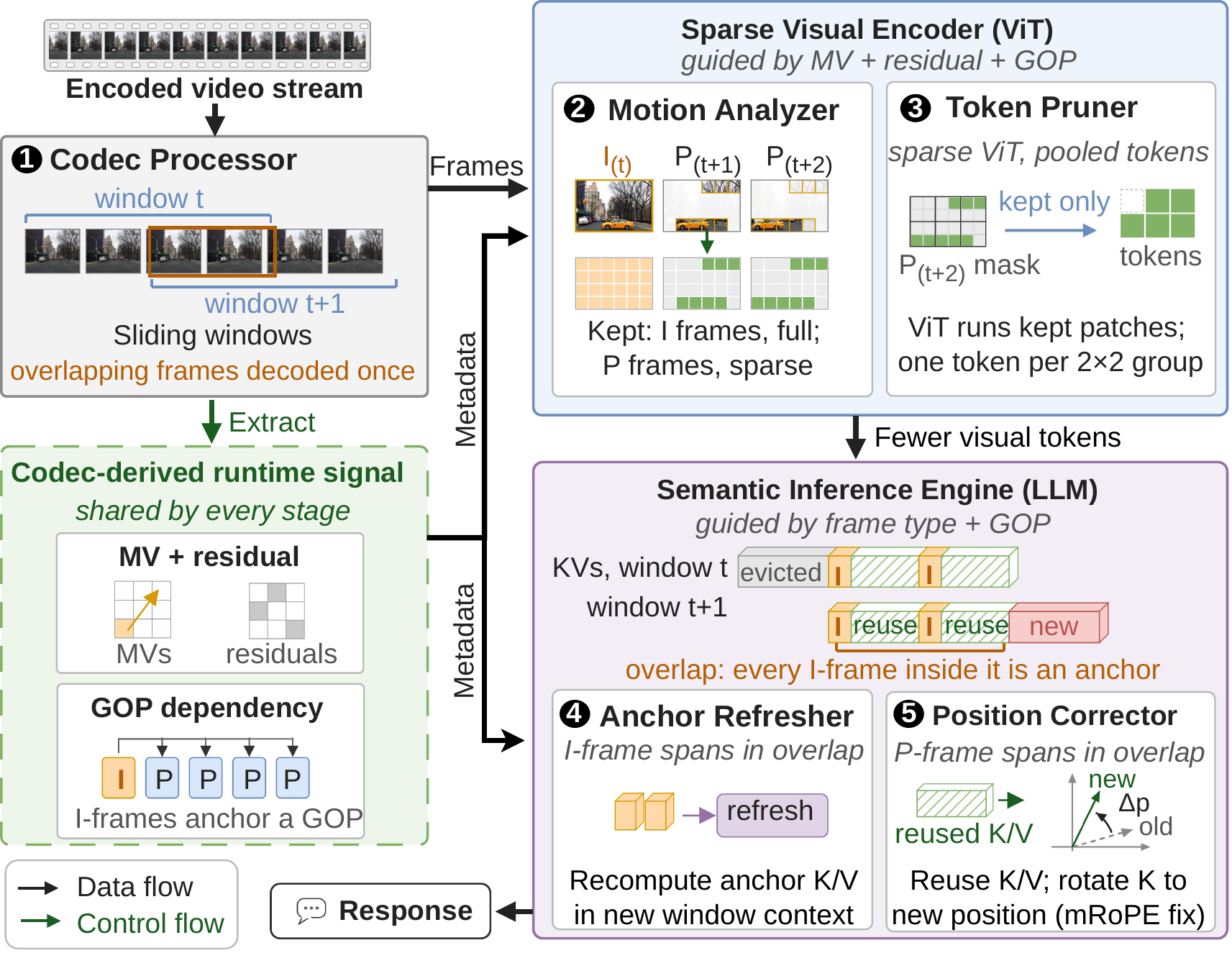}}
    \vspace{-1em}
    \caption{System architecture overview of \sys.}
    \vspace{-1em}
    \label{fig:costream_system}
\end{figure}

Fig.~\ref{fig:costream_system} shows the architecture of \sys. The \textbf{\textit{Codec Processor}}\ding{182} continuously decodes each compressed video stream and forms logical sliding windows over the decoded frames, allowing overlapping windows to reuse the same decoded frames. Along the same decode path, it extracts block-level motion/change information, frame types, and GOP boundaries as shared guidance for downstream stages.

To satisfy $R_1$, the \textbf{\textit{Motion Analyzer}}\ding{183} maps block-level motion/change signals onto the VLM patch layout and produces a patch mask for selective visual encoding. The \textbf{\textit{Token Pruner}}\ding{184} translates this mask into model-compatible sparse ViT execution, retaining complete projection groups when needed and producing fewer visual tokens for the LLM. Thus, sparsity identified before ViT execution reduces both visual computation and downstream LLM-prefill work.

To satisfy $R_2$, the \textit{Semantic Inference Engine} uses frame types and GOP boundaries to organize KV refresh and reuse, while maintaining correspondence among frames, visual tokens, and cached states across overlapping windows. The \textbf{\textit{Anchor Refresher}}\ding{185} recomputes KV states for I-frame token spans under the updated window context, while the \textbf{\textit{Position Corrector}}\ding{186} adjusts the positional encoding of cached keys for the remaining overlapping tokens. Anchor refresh and prefill for newly arrived tokens share the same execution path. Together, these mechanisms coordinate visual encoding and KV reuse
through shared codec metadata, without model-side redundancy analysis.

\vspace{-.5em}
\subsection{Hardware-Accelerated Codec Processing}
\label{sec:codec_processing}
\sys ingests each video stream as a compressed bitstream and decodes it using commodity GPU video engines~\cite{nvdec,ye2025sand,baobaid2025edge}, such as NVIDIA NVDEC. As the stream is decoded, the \textit{Codec Processor} extracts block-level motion information, frame types, and GOP boundaries and associates this metadata with the corresponding decoded frames. The metadata is thus available before ViT execution and guides both visual processing and cross-window KV reuse.

Decoding each sliding window independently would repeat work for frames shared by consecutive requests. \sys decouples sequential bitstream decoding from logical window formation: each stream is decoded once in arrival order into a per-stream temporal buffer, where every frame is stored together with its codec metadata. Logical windows are then constructed as views over this buffer, allowing overlapping requests to share decoded frames and their associated metadata. Decoded frames remain GPU-resident for preprocessing, where resizing, color-space conversion, and normalization are executed as batched GPU operations to avoid unnecessary CPU--GPU transfers. Throughout these operations, \sys explicitly tracks each frame's source identity and associated codec metadata, preserving their correspondence in the preprocessed model inputs.

\vspace{-.5em}
\subsection{Codec-Guided Selective Visual Encoding}
\label{sec:token_pruning}
To realize Opportunity\#1 and satisfy $R_1$, \sys uses codec-derived change signals to select patches before ViT execution. Prior work has established the utility of codec metadata for inter-frame reuse and visual-token reduction~\cite{song2024cmc,hwang2025dejavu,wu2026metom,sarkar2026cope,yang2026mage}. Turning these signals into model-compatible sparse execution requires mapping codec blocks onto each model's patch grid and respecting its visual-token projection layout. This enables \sys to reduce both ViT computation and the number of visual tokens entering LLM prefilling. The key challenge is to preserve query-relevant visual evidence while meeting these model-specific execution constraints.

\subsubsection{Codec Change Signal Analysis}
\label{sec:motion_vector_analysis}

\sys uses motion vectors as the default signal for identifying local visual changes, and optionally supplements them with residuals when tighter quality targets are required. Within a P-frame, motion vectors specify the spatial displacement from an inter-coded block to its reference region. For a coding block $m$ at time $t$, let $\mathbf{v}_t^m=(\Delta x,\Delta y)$ denote its motion vector. We use its magnitude as a block-level motion signal:
\vspace{-.5em}
\begin{equation}
    V_t^m = \|\mathbf{v}_t^m\|.
    \vspace{-.2em}
\end{equation}

Motion magnitude provides a lightweight signal for identifying regions that can potentially be skipped, but a small displacement does not rule out appearance changes within a block. Prediction residuals capture this remaining difference between the current block and its motion-compensated prediction~\cite{song2024cmc,sarkar2026cope,yang2026mage}. For block $m$, we quantify the residual using the sum of absolute differences:
\vspace{-.2em}
\begin{equation}
    R_t^m = \sum_{p \in B_t^m} \left|B_t^m(p)-\hat{B}_t^m(p)\right|,
    \vspace{-.5em}
\end{equation}
where $B_t^m$ and $\hat{B}_t^m$ denote the current block and its motion-compensated prediction, respectively. Thus, MVs capture spatial displacement, while residuals can reveal appearance changes that are not reflected by motion alone. 

\textit{Mapping codec changes to ViT patches.} Codec metadata and ViT execution use different spatial layouts: codec signals are defined over coding blocks in the source frame, whereas the ViT operates on a patch grid in the preprocessed model input. The \textit{Motion Analyzer} projects each coding block's spatial footprint onto this grid using the frame-to-input spatial scaling. Each patch receives the maximum motion-vector magnitude among blocks whose projected footprints overlap it, yielding the patch-level motion magnitude $V_t(i)$ for patch $i$.

By default, \sys uses $V_t(i)$ alone to determine candidate patches for pruning. When residual rescue is enabled, the \textit{Motion Analyzer} additionally maps residuals onto the same patch grid to obtain $R_t(i)$. Residuals affect the pruning decision only for patches that the motion-based policy would otherwise remove: patches with sufficiently large residuals are retained for ViT execution. Rescue therefore only adds patches to the retained set and never removes those already selected. The resulting patch mask is passed to the \textit{Token Pruner} for model-compatible sparse execution. \sys uses MV-only pruning by default and exposes residual rescue as an optional quality knob at additional cost. We evaluate this quality--cost trade-off in~\cref{sec:residual_rescue}.

\vspace{-.3em}
\subsubsection{Model-Compatible Token Pruning}
\label{sec:pruning_policy} Existing pruning methods~\cite{bolya2022token,rao2021dynamicvit} typically derive pruning or merging decisions from model-generated features or learned scores during inference. \sys instead derives patch-selection decisions from the codec signals mapped onto the ViT patch grid, before visual encoding begins, reducing both ViT computation and downstream visual tokens.

\sys retains all patches in I-frames, which provide no inter-frame motion vectors for motion-based selection. Keeping these frames intact preserves a complete visual representation at the start of each GOP. The following patch-selection rules apply to P-frames. Given the patch-level motion magnitude $V_t(i)$, \sys first marks patch $i$ as dynamic under the default MV-only policy:
\vspace{-.2em}
\begin{equation}
    D_t^{\mathrm{MV}}(i) = \mathbb{I}\!\left[V_t(i) \geq \tau_{\mathrm{MV}}\right],
    \vspace{-.2em}
\end{equation}
where $\tau_{\mathrm{MV}}$ controls the sensitivity of motion-based selection. A value of $D_t^{\mathrm{MV}}(i)=1$ marks patch $i$ for retention. Patches with $V_t(i)<\tau_{\mathrm{MV}}$ are candidates for pruning, but may still be retained by the additional rules below.

When residual rescue is enabled, patches not selected by the motion criterion are additionally retained if their residual magnitude meets or exceeds the threshold $\tau_{\mathrm{res}}$:
\vspace{-.3em}
\begin{equation}
    D_t^{\mathrm{res}}(i)
    = \mathbb{I}\!\left[
        D_t^{\mathrm{MV}}(i)=0
        \land R_t(i)\geq\tau_{\mathrm{res}}
      \right],
    \vspace{-.2em}
\end{equation}
with $D_t^{\mathrm{res}}(i)=0$ when residual rescue is disabled.

P-frames can also contain intra-coded regions, which are coded without reference to other frames. Patches overlapping these regions are retained because the absence of an inter-frame motion vector must not be interpreted as low motion. Let $D_t^{\mathrm{intra}}(i)$ indicate whether patch $i$ overlaps such a region. Combining these retention criteria gives the per-frame retention mask:
\vspace{-.3em}
\begin{equation}
    D_t(i)
    = D_t^{\mathrm{MV}}(i)
    \lor D_t^{\mathrm{res}}(i)
    \lor D_t^{\mathrm{intra}}(i).
    \vspace{-.2em}
\end{equation}

\begin{figure}[t]
\vspace{-.4em}
    \centerline{\includegraphics[width=.95\linewidth]{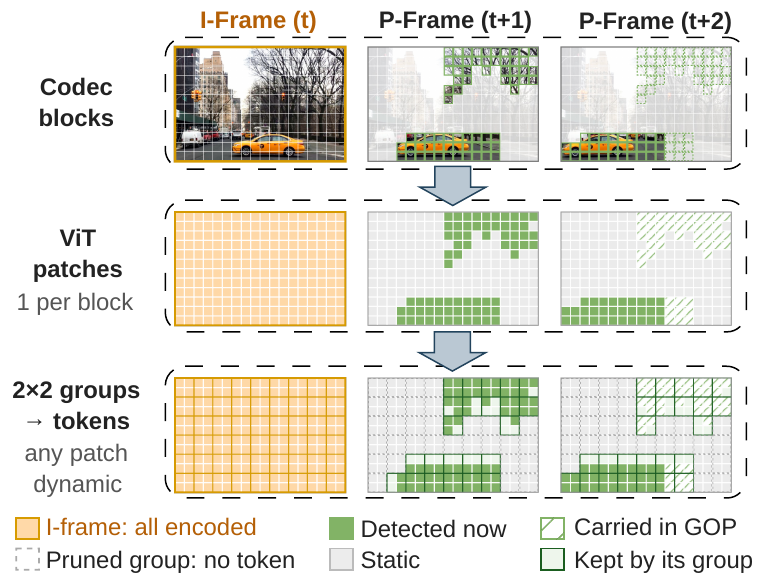}}
    \vspace{-1em}
    \caption{Illustration of the MV-guided pruning policy for a VLM with $(2\times2)$ spatial projection groups. A group is retained if any constituent patch is selected.}
    \vspace{-1.2em}
    \label{fig:pruning_policy}
\end{figure}

\textit{Temporal stabilization within a GOP.}
A per-frame change signal can miss regions whose measured change becomes temporarily small after an earlier event. \sys therefore accumulates the per-frame retention masks within each GOP. For each P-frame, the accumulated mask retains patch locations selected in the current frame or any preceding P-frame since the most recent I-frame. Once selected, a location remains active until the next I-frame resets the accumulation. The I-frame's full retention mask is excluded from this accumulation. This conservative policy preserves previously selected regions even when subsequent change signals become small, while bounding accumulation within each GOP. Fig.~\ref{fig:pruning_policy} illustrates this process.

\textit{Adapting sparsity to the model layout.}
The accumulated patch mask must then be adapted to the VLM's native visual-token projection layout. Many VLMs combine groups of neighboring ViT patches into fewer visual tokens through spatial merging, pixel shuffle, or pooling. \sys maps the patch mask through this grouping geometry: an output visual token is retained if any contributing patch is selected. The layout adapter maintains the correspondence between selected patches and output tokens during sparse visual execution. Selected embeddings retain their original relative order in the visual-token sequence, and only retained visual tokens are passed to the LLM.

The pruning policy remains common across VLMs; model-specific handling is confined to the layout adapter that maps the patch mask onto each model's patch grouping and projection structure. The same policy applies unchanged across the different visual layouts of all VLMs evaluated in~\cref{sec:evaluation}. This isolates architectural differences from the codec-derived selection and GOP-level mask policy.

\vspace{-.5em}
\subsection{Selective State Refresh}
\label{sec:selective_kvc_refresh}
To realize Opportunity~\#2 and satisfy $R_2$, \sys reduces repeated LLM prefilling across overlapping windows through selective KV refresh and reuse. KV reuse after visual pruning requires maintaining correspondence between retained tokens and cached states. Even when the retained tokens are unchanged, window advancement changes their positions and preceding context, so the cached states differ from those computed for the current window. Recent KV reuse methods address non-prefix or streaming reuse through selective recomputation, hierarchical state management, or position-invariant caches~\cite{yao2025cacheblend,qin2025vlcache,zhang2026hermes,ma2026kamera}. \sys uses frame types and GOP boundaries to guide selective refresh, while applying positional correction to reused keys.

To maintain token correspondence across windows, \sys derives visual selection from source-frame codec metadata and the accumulated history within each GOP, rather than window-dependent model representations. Under fixed preprocessing and pruning settings, this preserves the selected token layout of overlapping frames. The runtime associates cached states with both source-frame identity and the resulting token layout, ensuring that reuse operates on the corresponding retained tokens.

\begin{figure}[t] 
    \vspace{-.5em}
    \centerline{\includegraphics[width=1\linewidth]{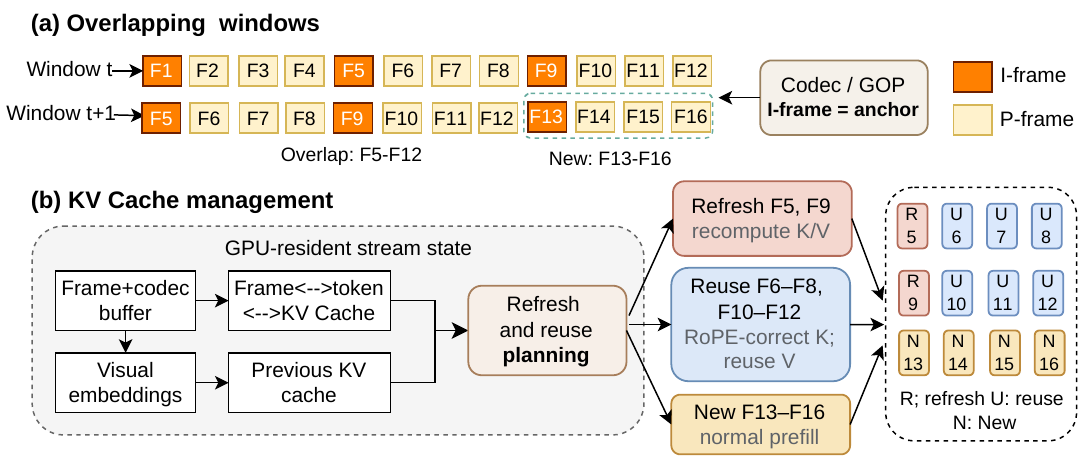}} 
    \vspace{-1em}
    \caption{Codec-guided KV management across overlapping windows.}
    \vspace{-1.3em}
    \label{fig:selective_kvc_refresh} 
\end{figure}

\vspace{-.3em}
\subsubsection{GOP-Guided Anchor Refresh}
\label{sec:kvc_refresh}

Fig.~\ref{fig:selective_kvc_refresh} illustrates an example stream with an I-frame every four frames. When a 12-frame window advances by four frames, $F5\!\sim\!F12$ remain in the overlap, accounting for 67\% of the frames in the new window. The cached visual embeddings of the overlapping frames can be reused, but their KV states require additional handling under the changed window context.

\sys uses frame types and GOP boundaries to select refresh anchors within the overlap. Visual-token spans corresponding to I-frames serve as anchors: these frames are retained in full by the visual pruning policy, providing complete-frame visual representations at identifiable points within each GOP. The \textit{Anchor Refresher} recomputes their KV states from cached visual embeddings under the current window context. In the example, it refreshes the spans for $F5$ and $F9$, while cached states for $F6\!-\!F8$ and $F10\!-\!F12$ are selected for reuse, with position correction described next. Tokens from newly arrived frames are prefilled normally.

The resulting refresh plan distinguishes refreshed anchors, reused overlapping states, and newly computed states. Together, selective refresh and reuse approximate the KV states produced by full recomputation while avoiding prefill over all overlapping tokens. We evaluate the quality and cost of this anchor policy in~\cref{sec:effectiveness_gop_refresh}.

\vspace{-.3em}
\subsubsection{Position-Corrected KV Reuse}
For overlapping states selected for reuse, the \textit{Position Corrector} adjusts cached keys to their positions in the current window. For a reused token $j$, let $\mathbf{p}_{\text{old}}(j)$ and $\mathbf{p}_{\text{new}}(j)$ denote its previous and current positional indices. For models with rotary position embeddings (RoPE)~\cite{su2024roformer,ma2026kamera}, \sys removes the rotation associated with the old position and applies the rotation for the current position:
\vspace{-0.2em}
    \begin{equation}
    \label{eq:rope_correction}
    \hat{K}_t(j) =
    R\!\left(\mathbf{p}_{\text{new}}(j)\right)
    R\!\left(\mathbf{p}_{\text{old}}(j)\right)^{-1}
    K_{t-1}(j),
    \vspace{-0.2em}
    \end{equation}
where $R(\cdot)$ denotes the model's RoPE rotation operator. For standard one-dimensional RoPE with fixed frequencies, the two rotations combine into $R(p_{\text{new}}(j)-p_{\text{old}}(j))$. For multidimensional RoPE, the correction follows the model's mapping from positional components to rotary dimensions.

Value states are retained without positional rotation, $\hat{V}_t(j)=V_{t-1}(j)$, since RoPE rotates queries and keys rather than values. Position correction updates the positional encoding but retains the representations computed under the previous context. Selective refresh and reuse therefore remain an approximation to full recomputation. We evaluate their task-quality impact and execution cost in~\cref{sec:effectiveness_gop_refresh}.

\vspace{-.5em}
\subsection{Multi-Stream Serving Runtime}
\label{sec:multistream}
In a multi-stream deployment, windows arrive asynchronously as individual video streams advance. \sys maintains independent per-stream state, including decoded frames, associated metadata, and reusable KV states. Windows from the same stream are processed in temporal order to preserve cross-window state reuse, while windows from different streams can be scheduled concurrently.

The runtime separates token-computation planning from KV-context preparation. GOP-guided anchor selection identifies refresh spans before LLM execution, without a separate model-side scoring pass. Only refresh tokens and the suffix of newly arrived tokens execute the Transformer; other overlapping tokens supply cached K/V as attention context. The runtime prepares this context by repacking reusable states into the current request's destination KV cache on the GPU.

The runtime brings these two paths together in a joint prefill pass for refresh tokens and the normal suffix. At each layer, their newly computed K/V are written directly into the destination cache before causal attention is evaluated. Attention therefore reads a combination of reused states and states computed in the current pass, allowing suffix tokens to attend to refreshed anchors. This integration avoids a separate refresh forward pass and an additional refresh-specific KV-cache copy. Refresh work also shares the normal prefill scheduling path across streams.

\vspace{-.5em}
\section{Implementation Details}
\label{sec:implementation}
We implement \sys on top of vLLM~\cite{vllm}, with 6.1K lines of Python code for codec processing, sparse visual execution, and cross-window KV-cache management.

\textbf{\textit{Codec Processing.}} The codec-processing front end consists of 600 lines of Python code and supports compressed H.264 streams~\cite{wiegand2003overview}. We use NVIDIA NVDEC via PyNvVideoCodec for hardware-accelerated decoding and per-block motion-vector extraction, while retaining frame/GOP metadata from the codec stream. Since NVDEC does not expose prediction residuals through the same interface, we compute a reconstruction-error proxy on the GPU using decoded frames and motion vectors. Each patch's residual score is the mean absolute difference from the motion-compensated prediction of the previous decoded frame. This path is optional; our default MV-only setup relies solely on NVDEC and incurs no residual-extraction overhead.

\textbf{\textit{Sparse Visual Encoding.}} We integrate codec-guided sparse encoding into vLLM's visual-encoder path with approximately 3,000 lines of Python code. Lightweight model-specific adapters map a shared pruning mask to each VLM's native patch-grouping and visual-projection layout, preserving a common pruning policy across models.

\textbf{\textit{Selective State Refresh.}} We integrate selective refresh into vLLM's scheduler and worker with approximately 2,500 lines of Python code. A frame-aware index maps visual-token spans to physical KV slots across cache-block boundaries, while KV tensors remain in vLLM's GPU block pool. Position-handling routines support standard RoPE and Qwen3-VL's three-axis mRoPE.

\vspace{-.5em}
\section{Methodology}
\label{sec:experimental_setup}

\textbf{Testbed and models.}
We evaluate three VLM families with different visual encoders and LLM scales (Table~\ref{tab:models_configurations}). Experiments run on servers with AMD EPYC 7713 64-Core, 512\,GB host memory, CUDA 13.1, and NVIDIA driver 590.48.01. LLaVA-OneVision runs on one 80\,GB A100, InternVL3 on one 94\,GB H100, and Qwen3-VL on two 94\,GB H100 GPUs with tensor parallelism via PCIe 4.0 (TP=2).

\begin{table}[t]
\vspace{-.5em}
\centering
\caption{Models and deployment configurations.}
\label{tab:models_configurations}
\vspace{-1em}
\resizebox{\linewidth}{!}{
\begin{tabular}{c|c|c|c}
\toprule
\textbf{Model} & \textbf{ViT Encoder} & \textbf{LLM Backbone} & \textbf{Deployment} \\
\midrule
LLaVA-OneVision~\cite{li2024llavaov}
    & SigLIP ViT-SO (400M) & Qwen2-7B & 1$\times$A100 (80\,GB) \\
\hline
InternVL3~\cite{zhu2025internvl3}
    & InternViT (300M) & Qwen2.5-14B & 1$\times$H100 (94\,GB) \\
\hline
Qwen3-VL~\cite{bai2025qwen3}
    & Qwen-ViT (600M) & Qwen3-32B & 2$\times$H100 (188\,GB) \\
\bottomrule
\end{tabular}
}
\begin{minipage}{\linewidth}
\vspace{.2em}
\footnotesize\textit{Note:} LLaVA-OneVision is abbreviated as LLaVA-OV. Qwen3-VL uses TP=2.
\end{minipage}
\vspace{-1.7em}
\end{table}

\textbf{Baselines.}
We compare \sys against four representative baselines. \textit{\textbf{Full Compute}} (Full Comp) uses vLLM~\cite{vllm} to process all sampled frames and visual tokens in every window without pruning or cross-window KV reuse. \textit{\textbf{Déjà Vu}}~\cite{hwang2025dejavu} exploits inter-frame similarity to reuse visual computation and reduce ViT cost. \textit{\textbf{CacheBlend}}~\cite{yao2025cacheblend} reuses non-prefix cached states and selectively recomputes a top-$k$ subset of tokens to recover context consistency. \textit{\textbf{VLCache}}~\cite{qin2025vlcache} jointly reuses visual features and KV states with a dynamic layer-aware recomputation policy.

For mechanism analysis, we also compare with \textit{\textbf{FastV}}~\cite{chen2024image}, which prunes tokens by attention importance, and \textit{\textbf{FrameFusion}}~\cite{fu2025framefusion}, which uses similarity and importance. Both derive pruning signals during model execution.

\textbf{Workloads and request protocol.}
We replay all four workloads as streaming video under a common sliding-window protocol. \textit{\textbf{UCF-Crime}}~\cite{sultani2018real} and \textit{\textbf{XD-Violence}}~\cite{wu2020not} cover anomaly and violence detection, while \textit{\textbf{StreamingBench}}~\cite{lin2026streamingbench} and \textit{\textbf{OVO-Bench}}~\cite{niu2025ovo} cover streaming video Q\&A. For the latter two, we retain all questions associated with real-time tasks. For controlled evaluation, videos are re-encoded with an 8\,s GOP interval; inference samples frames at 2\,FPS. Each request contains a 40\,s window (80 frames), and windows advance by 8\,s (16 frames).

\textbf{Metrics.}
For inference quality, we report frame-level ROC-AUC for UCF-Crime and XD-Violence, and exact-choice accuracy for StreamingBench and OVO-Bench. For serving performance, we report mean TTFT and throughput, measured as completed windows per second. With each stream submitting one window every 8\,s, $N$ streams impose an offered load of $0.125N$ requests/s. A stream count is \emph{sustainable} if throughput reaches at least 95\% of the offered load over 64\,s, with no sustained backlog growth or failures.

\vspace{-1.5em}
\section{Evaluation}
\label{sec:evaluation}
We evaluate \sys across models and workloads for quality--latency trade-offs, scalability, resource efficiency, cross-stage mechanisms, and parameter sensitivity. Unless otherwise specified, we use a 40\,s window; the 8\,s stride, 16-frame GOP, and MV threshold of 0.5 are selected from the sweeps in~\cref{sec:sensitivity} for the best quality--latency trade-off.

\begin{figure}[t]
\vspace{-.5em}
    \centering
    \includegraphics[width=\linewidth]{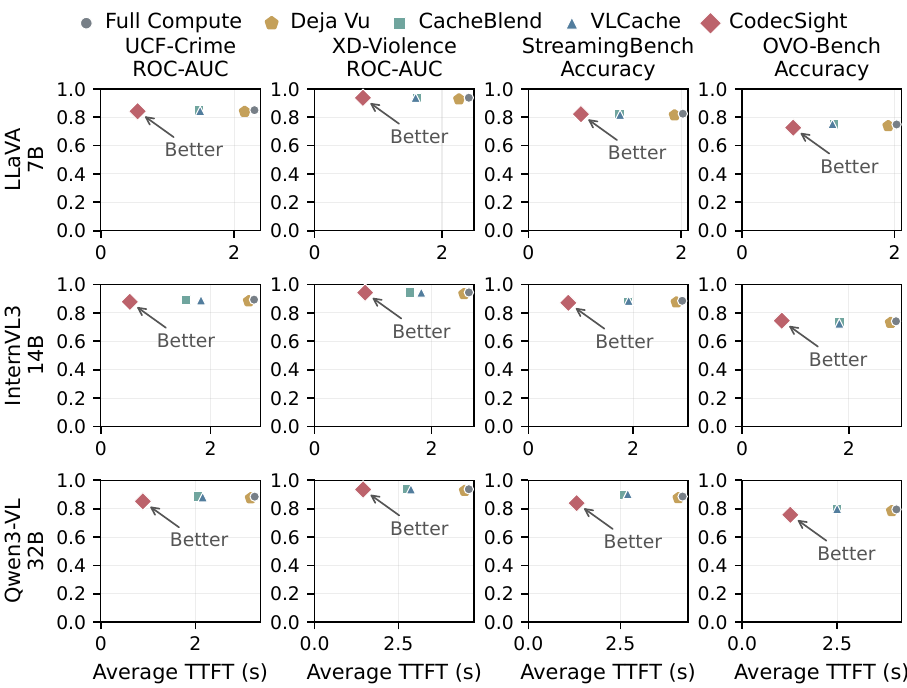}
    \vspace{-2em}
    \caption{ End-to-end quality--latency trade-off across three VLMs and four workloads. 
    }
    \label{fig:e2e_accuracy_ttft}
    \vspace{-1em}
\end{figure}

\vspace{-.3em}
\subsection{End-to-End Performance}
\label{sec:e2e_performance}

\subsubsection{Quality--latency trade-off.}
\label{sec:e2e_latency_quality}
Fig.~\ref{fig:e2e_accuracy_ttft} summarizes all 12 model--workload combinations. Compared with \textit{Full Comp}, \sys achieves a $2.94$$\sim$$5.28\times$ TTFT speedup, averaging $3.55\times$. It is also $1.73$$\sim$$2.94\times$ faster than CacheBlend and VLCache, consistently achieving the lowest TTFT. The advantage comes from reducing redundancy across stages. D\'ej\`a Vu targets visual-side reuse, but ViT encoding accounts for only $5$$\sim$$12\%$ of \textit{Full Comp} TTFT, limiting its end-to-end benefit. CacheBlend and VLCache reduce repeated state computation without pre-ViT sparsity. In contrast, \sys prunes before ViT execution, shortens the visual-token sequence entering LLM prefilling, and reuses overlapping KV states across windows, reducing both new visual/prefill work and repeated prefill over the overlap.

These savings retain most task quality. On frame-level ROC-AUC benchmarks, the gap from \textit{Full Comp} ranges from 0.07 to 3.37 percentage points (p.p.). \sys remains within 0.21 p.p. on XD-Violence across three models, while UCF-Crime gaps range from 0.77 to 3.37 p.p. On video-QA, exact-choice accuracy changes from a 0.38-point improvement to a 4.64-p.p. decrease. LLaVA-OV and InternVL3 remain within 1.43 points on StreamingBench, while Qwen3-VL shows the largest reductions on StreamingBench and OVO-Bench. Overall, \sys delivers substantial latency savings while retaining quality across anomaly-detection and video-QA workloads.

\begin{figure}[t]
    \vspace{-.5em}
    \centering
    \includegraphics[width=\linewidth]{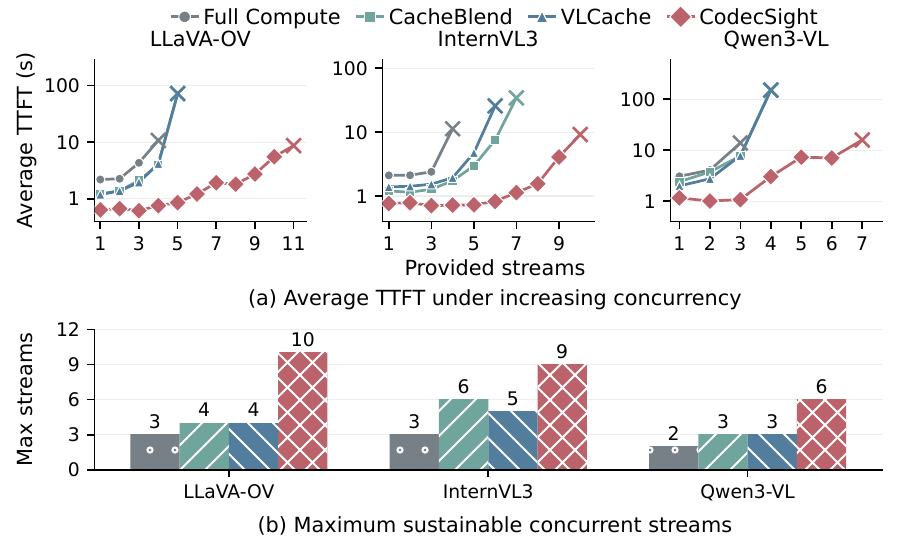}
    \vspace{-2em}
    \caption{Multi-stream scalability on UCF-Crime. (a) Mean TTFT with increasing concurrency; crosses indicate unsustainable stream counts. (b) Maximum sustainable streams.}
    \label{fig:e2e_throughput}
    \vspace{-0.8em}
\end{figure}

\vspace{-.3em}
\subsubsection{Multi-stream scalability.}
\label{sec:e2e_throughput}
We next evaluate whether \sys's per-request savings persist under concurrent streaming. At $N{=}1$, \sys reduces mean TTFT to 0.64, 0.76, and 1.16\,s for LLaVA-OV, InternVL3, and Qwen3-VL, respectively, a $2.7$$\sim$$3.4\times$ speedup over \textit{Full Comp}. This advantage persists as concurrency increases (Fig.~\ref{fig:e2e_throughput}(a)). \sys sustains 10, 9, and 6 streams on the three models, versus 3, 3, and 2 for \textit{Full Comp} (Fig.~\ref{fig:e2e_throughput}(b)), improving sustainable capacity by up to $3.3\times$. Relative to the strongest cache baseline, \sys supports $2.5\times$, $1.5\times$, and $2.0\times$ more streams, respectively. Cache-only reuse reduces repeated LLM computation but leaves the visual-token workload largely unchanged. In contrast, \sys reduces the visual and KV working set through pre-ViT pruning, while selective KV refresh removes repeated prefill over overlapping windows. Together, these savings reduce compute and memory pressure, enabling the highest concurrency across all three deployments.

\begin{figure}[t]
\vspace{-.7em}
    \centering
    \includegraphics[width=\linewidth]{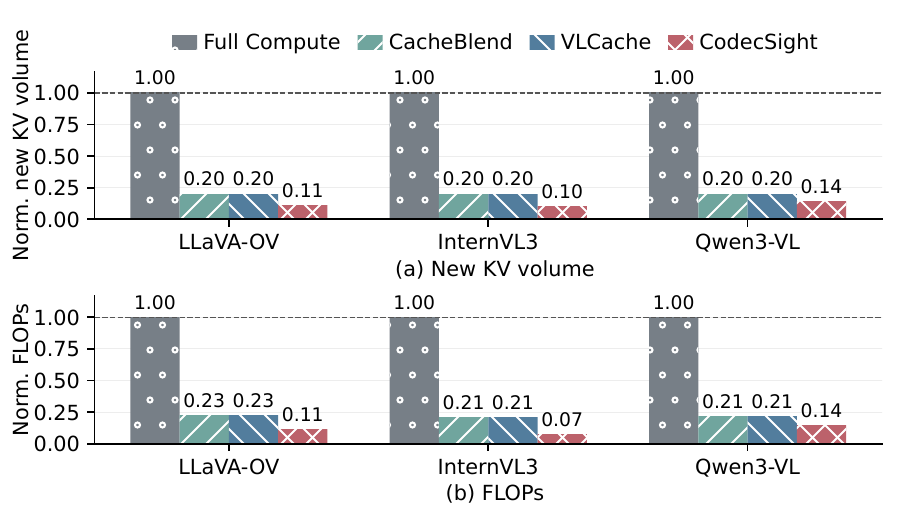}
    \vspace{-2.2em}
    \caption{Resource demand normalized to \textit{Full Comp} on UCF-Crime. (a) New KV volume per steady window. (b) Executed FLOPs under single-stream serving.}
    \vspace{-1em}
    \label{fig:resource_savings}
\end{figure}

\vspace{-.3em}
\subsubsection{Resource efficiency}
\label{sec:resource_savings}
The scalability gains stem from reducing both new KV-state generation and executed computation (Fig.~\ref{fig:resource_savings}). With a 40\,s window and an 8\,s stride, 20\% of each steady window contains newly arrived frames. Accordingly, CacheBlend and VLCache require 0.20 of \textit{Full Comp}'s new KV volume under full-overlap-hit accounting, since states must still be created for the new content. \sys further prunes the newly arrived GOP before ViT execution, so only retained visual tokens generate downstream KV states. This reduces new KV volume to 0.10$\sim$0.14 of \textit{Full Comp}, an 86$\sim$90\% reduction and a further 29$\sim$48\% reduction over the cache baselines.

The same cross-stage effect appears in computation. CacheBlend and VLCache execute 0.21$\sim$0.23 of \textit{Full Comp}'s FLOPs: they reduce cached-token recomputation, but these savings are not coordinated with pre-ViT sparsity. In contrast, \sys combines sparse visual execution with cross-window KV reuse, so pruning reduces ViT work and the visual-token sequence entering the LLM, while KV reuse removes repeated prefill over the overlap. \sys therefore executes 0.07$\sim$0.14 of \textit{Full Comp}'s FLOPs, an 86$\sim$93\% reduction and 34$\sim$66\% less than CacheBlend and VLCache.

\vspace{-.3em}
\subsection{Mechanism and Policy Analysis}
\label{sec:each_component}
Unless otherwise specified, all experiments in this subsection use InternVL3 on UCF-Crime.

\begin{figure}[t]
    \centering
    \begin{subfigure}[t]{.48\linewidth}
        \centering
        \includegraphics[width=\linewidth]{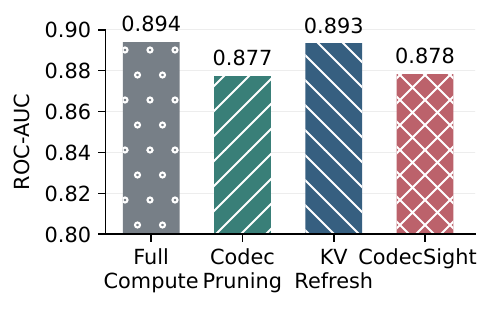}
        \vspace{-2em}
        \caption{Inference quality}
        \label{fig:ablation_accuracy}
    \end{subfigure}
    \hfill
    \begin{subfigure}[t]{.48\linewidth}
        \centering
        \includegraphics[width=\linewidth]{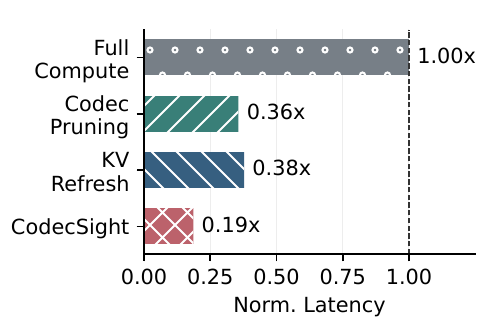}
        \vspace{-2em}
        \caption{Normalized latency}
        \label{fig:ablation_latency}
    \end{subfigure}
    \vspace{-1em}
    \caption{Ablation of \sys's cross-stage mechanisms: codec pruning, selective KV refresh, and their combination.}
    \label{fig:formal_ablation}
    \vspace{-1em}
\end{figure}

\vspace{-.3em}
\subsubsection{Cross-stage mechanism ablation}
Fig.~\ref{fig:formal_ablation} isolates \sys's two mechanisms. Codec Pruning reduces normalized TTFT to 0.36 with 0.877 ROC-AUC, showing that visual sparsity lowers serving cost but drives the quality trade-off. KV Refresh reaches 0.38 normalized TTFT with 0.893 ROC-AUC, close to \textit{Full Comp}'s 0.894, indicating that overlap reuse removes repeated prefilling with little quality impact.
Combining both reduces normalized TTFT to 0.19 while maintaining 0.878 ROC-AUC. Quality remains unchanged from \textit{Codec Pruning}, showing that \textit{KV Refresh} does not compound the pruning-induced quality loss. Combined latency is 47\% lower than \textit{Codec Pruning} and 50\% lower than \textit{KV Refresh}. The mechanisms provide complementary savings: codec-guided pruning reduces ViT computation and the visual-token sequence entering LLM prefilling, while overlap reuse and selective refresh eliminate repeated prefill for processed content.

\begin{figure}[t]
\vspace{-.5em}
    \centering
    \begin{subfigure}[t]{.48\linewidth}
        \centering
        \includegraphics[width=\linewidth]{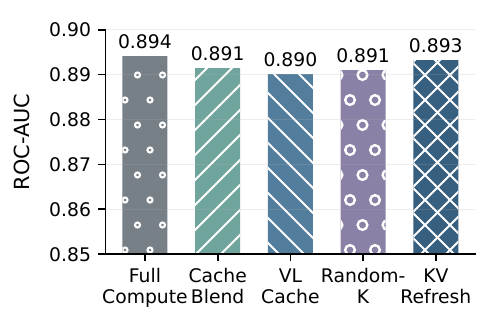}
        \vspace{-2em}
        \caption{Inference quality}
        \label{fig:kv_refresh_accuracy}
    \end{subfigure}
    \hfill
    \begin{subfigure}[t]{.48\linewidth}
        \centering
        \includegraphics[width=\linewidth]{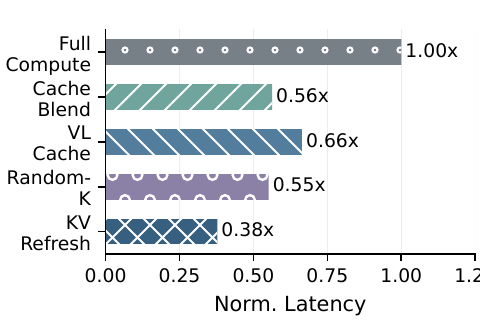}
        \vspace{-2em}
        \caption{Normalized latency}
        \label{fig:kv_refresh_latency}
    \end{subfigure}
    \vspace{-1em}
    \caption{Effectiveness of GOP-guided selective refresh under a matched $1/16$ recomputation budget.}
    \label{fig:kv_refresh_policy}
    \vspace{-1em}
\end{figure}

\vspace{-.3em}
\subsubsection{Effectiveness of GOP-guided refresh.}
\label{sec:effectiveness_gop_refresh}
Fig.~\ref{fig:kv_refresh_policy} evaluates both the refresh policy and its execution mechanism. Under the same $1/16$ recomputation budget, GOP-guided refresh achieves 0.893 ROC-AUC, close to \textit{Full Comp}'s 0.894 and higher than CacheBlend (0.891), VLCache (0.890), and Random-$K$ (0.891). This shows that codec-defined anchors better identify where limited recomputation should be spent as the window context changes.
\sys also achieves the lowest normalized TTFT at 0.38, compared with 0.55$\sim$0.66 for the selective-refresh baselines. Beyond lightweight GOP-based planning, \sys introduces an in-place refresh path that recomputes selected anchor states directly into the GPU-resident KV cache, avoiding the separate cache materialization used by the baselines. This design reduces KV-management overhead while preserving the quality benefit of selective recomputation.

\begin{figure}[t]
    \centering
    \begin{subfigure}[t]{.48\linewidth}
        \centering
        \includegraphics[width=\linewidth]{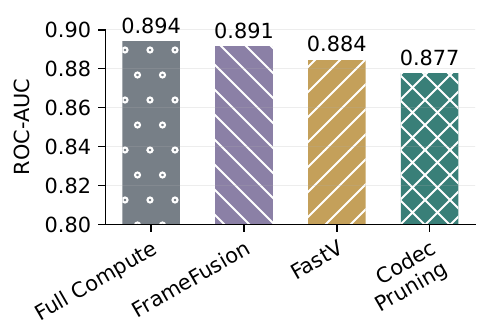}
        \vspace{-2em}
        \caption{Inference quality}
        \label{fig:pruning_policy_accuracy}
    \end{subfigure}
    \hfill
    \begin{subfigure}[t]{.48\linewidth}
        \centering
        \includegraphics[width=\linewidth]{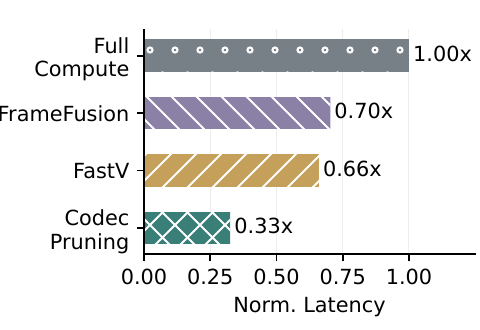}
        \vspace{-2em}
        \caption{Normalized latency}
        \label{fig:pruning_policy_latency}
    \end{subfigure}
    \vspace{-1em}
    \caption{\sys's Codec pruning versus FastV and FrameFusion using the same 50\% visual-token budget.}
    \vspace{-0.8em}
    \label{fig:pruning_policy_comparison}
\end{figure}

\vspace{-.3em}
\subsubsection{Benefit of Codec pruning}
As shown in Fig.~\ref{fig:pruning_policy_comparison}, at the matched 50\% visual-token budget, FastV and FrameFusion achieve 0.884 and 0.891 ROC-AUC, respectively, compared with 0.877 for Codec Pruning. Their normalized TTFT, however, remains at 0.66 and 0.70, whereas Codec Pruning reduces it to 0.33.
The difference stems from when sparsity becomes available. FastV and FrameFusion derive pruning signals during model execution, so part of the computation required to obtain these signals has already been incurred before tokens are removed. In contrast, \sys derives pruning decisions from codec metadata before ViT execution, allowing pruned patches to bypass visual encoding and preventing their visual tokens from entering LLM prefilling. This early sparsity reduces normalized TTFT by roughly $2.0$$\sim$$2.1\times$ relative to the two model-side policies, at a 0.7$\sim$1.4 percentage-point ROC-AUC drop.

\begin{figure}[t]
\vspace{-.5em}
    \centering
    \begin{subfigure}[t]{.48\linewidth}
        \centering
        \includegraphics[width=\linewidth]{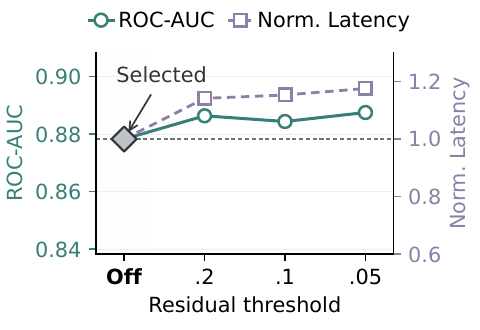}
        \vspace{-2em}
        \caption{Quality--latency tradeoff}
        \label{fig:residual_tradeoff}
    \end{subfigure}
    \hfill
    \begin{subfigure}[t]{.48\linewidth}
        \centering
        \includegraphics[width=\linewidth]{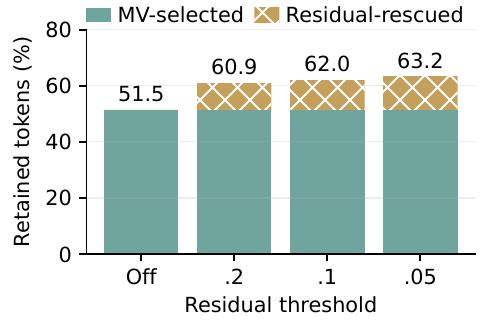}
        \vspace{-2em}
        \caption{Retained tokens}
        \label{fig:residual_tokens}
    \end{subfigure}
    \vspace{-1em}
    \caption{Effect of residual rescue at a fixed MV threshold of 0.5. TTFT is normalized to MV-only pruning.}
    \vspace{-1em}
    \label{fig:residual_rescue}
\end{figure}

\vspace{-.3em}
\subsubsection{Residual rescue as a quality knob.}
\label{sec:residual_rescue}
Motion vectors provide a lightweight change signal, but can miss appearance changes not captured by displacement. Residuals complement MVs by rescuing such regions at additional extraction and model-execution cost. As shown in Fig.~\ref{fig:residual_rescue}, a residual threshold of 0.2 raises ROC-AUC from 0.878 to 0.886 by rescuing 9.4\% of dense visual tokens, while increasing normalized TTFT to 1.14. Lowering the threshold to 0.05 retains 63.2\% of visual tokens and improves ROC-AUC to 0.888, but raises TTFT to 1.18.
The diminishing quality gain comes with increasing residual-extraction and retained-token overhead: residual rescue adds 14$\sim$18\% latency over MV-only pruning. We therefore use MV-only pruning by default and expose residual rescue as an optional quality knob for deployments that favor higher accuracy.

\vspace{-.3em}
\subsection{Sensitivity of Design Parameters}
\label{sec:sensitivity}

\begin{figure}[t]
    \centering
    \includegraphics[width=\linewidth]{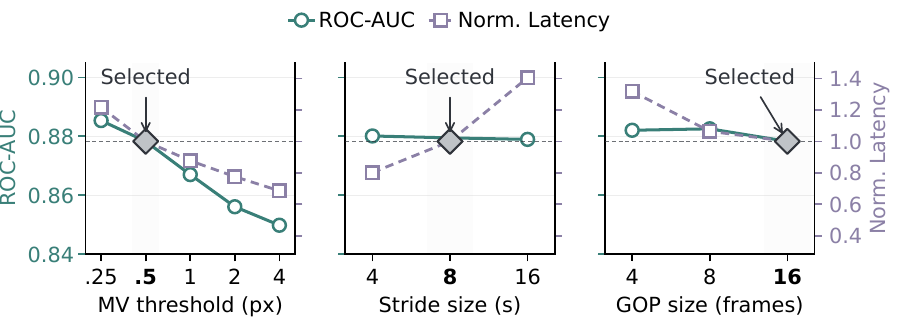}
    \vspace{-2.2em}
    \caption{Quality--latency trade-offs under different MV thresholds, stride sizes and GOP sizes.}
    \vspace{-1em}
    \label{fig:sensitivity_tradeoffs}
\end{figure}

\textbf{MV threshold.} The MV threshold controls pruning aggressiveness by determining how much low-motion content is treated as static. In Fig.~\ref{fig:sensitivity_tradeoffs}, increasing the threshold from 0.25 to 4 pixels reduces retained tokens from 68.7\% to 24.8\% and normalized TTFT from 1.26 to 0.68, while ROC-AUC decreases from 0.885 to 0.850. The selected threshold of 0.5 retains 51.5\% of visual tokens and achieves 0.878 ROC-AUC. Relative to 0.25, it reduces retained tokens by 17.2 p.p. and latency by 21\% for only a 0.71 p.p. ROC-AUC decrease; more aggressive thresholds incur substantially larger quality loss. This knee in the quality--latency curve motivates MV${=}0.5$ for the end-to-end evaluation.

\textbf{Stride size.} As shown in Fig.~\ref{fig:sensitivity_tradeoffs}, the stride controls the trade-off between update frequency, serving load, and cross-window reuse. With a fixed 40\,s window, reducing the stride from 8\,s to 4\,s improves ROC-AUC by only 0.06 p.p. and lowers per-request TTFT by 19.8\%, but doubles the request arrival rate. Increasing the stride to 8 or 16\,s reduces window overlap, with ROC-AUC decreasing from 0.88 to 0.879 and normalized TTFT rising from 1.00 to 1.40. We therefore use an 8\,s stride to balance update frequency, per-request latency, and request volume.

\textbf{GOP size.} GOP size determines the frequency of codec-defined I-frame anchors and affects both visual processing and cross-window KV refresh. In Fig.~\ref{fig:sensitivity_tradeoffs}, increasing the GOP size from 4 to 16 reduces normalized average TTFT from 1.32 to 1.0, with only a small ROC-AUC drop from roughly 0.882 to 0.878. Smaller GOPs introduce more frequent full I-frame processing and anchor KV recomputation, improving quality but increasing ViT and LLM-prefill cost. We therefore use GOP=16 by default for its favorable quality--latency trade-off, while keeping it configurable.

\begin{figure}[t]
\vspace{-.5em}
  \centering
  \begin{subfigure}[t]{.48\linewidth}
      \centering
      \includegraphics[width=\linewidth]{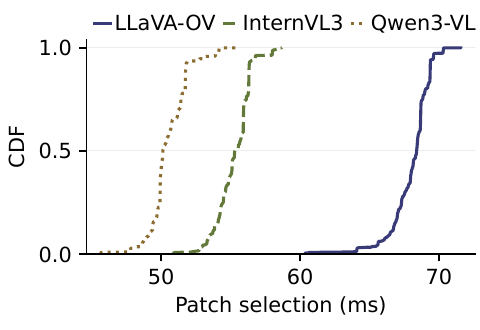}
      \vspace{-2em}
      \caption{Patch selection}
      \label{fig:token_selection_overhead}
  \end{subfigure}
  \hfill
  \begin{subfigure}[t]{.48\linewidth}
      \centering
      \includegraphics[width=\linewidth]{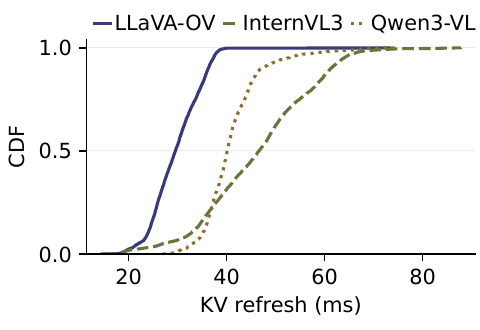}
      \vspace{-2em}
      \caption{KV refresh}
      \label{fig:kv_refresh_overhead}
  \end{subfigure}
  \vspace{-1.2em}
  \caption{CDFs of \sys' runtime overhead.}
  \vspace{-1em}
  \label{fig:system_overhead}
\end{figure}

\vspace{-.3em}
\subsection{Runtime Overhead}
As shown in Fig.~\ref{fig:system_overhead}, we measure the overhead of \sys's two online control paths: codec-guided patch selection and selective KV refresh. Patch selection incurs a median overhead of 50.1$\sim$68.4\,ms across the three models, accounting for 5.3$\sim$10.4\% of mean TTFT. KV refresh is cheaper, with a median overhead of 29.6$\sim$46.6\,ms, or 4.3$\sim$6.2\% of mean TTFT.
These costs remain small because patch decisions are derived directly from codec metadata without model-side scoring, while reusable KV states remain GPU-resident and are updated in place during refresh. The resulting control overhead is therefore substantially smaller than the computation removed by pre-ViT pruning and cross-window KV reuse, preserving \sys's end-to-end latency speedups.

\vspace{-.3em}
\subsection{Generalization Across Codecs}
\sys also extends to other codecs that expose motion information and frame dependencies. We validate this portability on H.265 using the same setup as~\cref{sec:each_component}, without modifying the pipeline. Compared with H.264, \sys on H.265 achieves comparable quality (0.865 vs. 0.878 ROC-AUC), retains fewer visual tokens (37.6\% vs. 51.5\%), and lowers mean TTFT by 17.9\%. These results demonstrate that \sys's cross-stage design generalizes beyond H.264 when the required codec metadata is available.

\vspace{-.3em}
\section{Related Work}
\label{sec:related_work}
\textbf{Codec-assisted video processing.}
Prior video analytics systems reduce processing cost through query optimization, approximation, cascades, and indexing~\cite{romero2022optimizing,zhang2017live,kang2017noscope,hsieh2018focus}. Related compressed-domain systems exploit codec or motion signals before pixel-domain processing. CoVA~\cite{hwang2022cova} reduces decoding and inference costs through compressed-domain analysis, Boggart~\cite{agarwal2023boggart} builds motion- and tracking-based indices for retrospective analytics, and SLICE~\cite{jung2026slice} uses motion vectors and residuals for selective video processing. These systems target front-end video processing or conventional vision inference. \sys instead uses codec metadata as a shared runtime signal to coordinate visual reduction and LLM state reuse across streaming VLM execution.

\textbf{Selective visual processing.}
Prior methods reduce visual computation by pruning, merging, or compressing redundant tokens using similarity, attention, or importance signals~\cite{bolya2022token,chen2024image,fu2025framefusion,zhangsparsevlm,xing2024pyramiddrop,ye2025fit,yang2025visionzip,zhang2025beyond,zhang2025vscan,tao2025dycoke,wang2026earlytom}. Video-oriented methods further exploit temporal redundancy through adaptive pruning, feature reuse, or streaming token compression~\cite{yao2025timechat,chen2025streamingtom,jiang2026stateful,regmi2026adaptive,wang2026stc}. Codec-guided methods including CMC~\cite{song2024cmc}, D\'ej\`a Vu~\cite{hwang2025dejavu}, COPE~\cite{sarkar2026cope}, MeToM~\cite{wu2026metom}, and Mage-VL~\cite{yang2026mage} exploit motion, residual, or metadata signals for sparse visual processing. \sys instead makes pruning decisions before ViT execution on pretrained VLMs, allowing the resulting sparsity to reduce both visual computation and downstream LLM prefilling.

\textbf{KV-Cache management for streaming VLMs.}
KV-cache management reduces long-context serving cost through paging and allocation~\cite{kwon2023efficient,prabhu2024vattention}, cache reuse~\cite{yao2025cacheblend,cheng2025lmcache,qin2025vlcache,ma2026kamera}, and eviction, compression, or quantization~\cite{zhang2023h2o,li2024snapkv,cai2024pyramidkv,liu2024kivi}. Streaming-video systems manage visual context through cache construction, compression, retrieval, or hierarchical memory, including DSCache~\cite{pang2026decouple}, StateKV~\cite{eyzaguirre2026linear}, StreamKV~\cite{chen2026streamkv}, HERMES~\cite{zhang2026hermes}, MuKV~\cite{xiao2026mukv}, and V-Rex~\cite{kim2026v}. These approaches manage growing or retrieved history from model-side state. \sys instead targets repeated inference over overlapping windows, using codec frame/GOP structure to selectively refresh context-sensitive states while reusing overlap, thereby reducing LLM prefilling.

\vspace{-.3em}
\section{Conclusion}
This paper presents \sys, a codec-guided serving system for streaming video understanding. \sys exposes codec-derived motion and GOP dependencies as shared runtime signals to coordinate pre-ViT visual sparsity with cross-window KV refresh. By reducing redundant computation before it propagates through the VLM pipeline, \sys improves multi-stream scalability and serving latency with low runtime overhead while preserving task quality.

\begin{acks}
The authors thank the members of the HyScale lab at NTU Singapore for their constructive discussions and feedback on this work. This project is supported by the Ministry of Education, Singapore, under its Academic Research Funds Tier 1 RG110/25 and RS26/23, and A*STAR Graduate Scholarship.
\end{acks}

\newpage
\balance
\bibliographystyle{ACM-Reference-Format}
\bibliography{bibcloud/gen-abbrev,dblp,misc}

\sloppypar
\end{document}